\documentclass[acmsmall,screen]{acmart}
\usepackage{booktabs} 
\usepackage{pifont}
\usepackage{multirow}
\usepackage{subfigure}
\usepackage{array}
\usepackage{booktabs}
\usepackage{float}
\usepackage{hyperref}
\usepackage{setspace}
\usepackage{threeparttable}
\usepackage{supertabular}
\usepackage{bm}
\usepackage{amsmath}
\usepackage{amsthm}
\usepackage{mathrsfs}

\newcommand{\newadd}[1]{{\color{black}#1}}
\newcommand{\qa}[1]{{\color{black}#1}}
\AtBeginDocument{%
  \providecommand\BibTeX{{%
    \normalfont B\kern-0.5em{\scshape i\kern-0.25em b}\kern-0.8em\TeX}}}

\begin{document}

\title{Caseformer: Pre-training for Legal Case Retrieval Based on Inter-Case Distinctions}


\author{Weihang Su}
\email{swh22@mails.tsinghua.edu.cn}
\affiliation{%
  \institution{Quan Cheng Laboratory, Department of Computer Science and Technology, Institute for Internet Judiciary, Tsinghua University}
  \city{Beijing}
  \country{China}
}

\author{Qingyao Ai}
\email{aiqy@tsinghua.edu.cn}
\authornote{Corresponding author}
\affiliation{%
  \institution{Department of Computer Science and Technology, Institute for Internet Judiciary, Tsinghua University}
  \city{Beijing}
  \country{China}
}

\author{Yueyue Wu}
\email{wuyueyue1600@gmail.com}
\affiliation{%
  \institution{Department of Computer Science and Technology, Institute for Internet Judiciary, Tsinghua University}
  \city{Beijing}
  \country{China}
}

\author{Yixiao Ma}
\email{ma-yx16@tsinghua.org.cn}
\affiliation{%
  \institution{Department of Computer Science and Technology, Institute for Internet Judiciary, Tsinghua University}
  \city{Beijing}
  \country{China}
}

\author{Haitao Li}
\email{liht22@mails.tsinghua.edu.cn}
\affiliation{%
  \institution{Department of Computer Science and Technology, Institute for Internet Judiciary, Tsinghua University}
  \city{Beijing}
  \country{China}
}

\author{Zhijing Wu}
\email{zhijingwu@bit.edu.cn}
\affiliation{%
  \institution{School of Computer Science and Technology, Beijing Institute of Technology}
  \city{Beijing}
  \country{China}
}

\author{Yiqun Liu}
\email{yiqunliu@tsinghua.edu.cn}
\affiliation{%
  \institution{Department of Computer Science and Technology, Institute for Internet Judiciary, Tsinghua University}
  \city{Beijing}
  \country{China}
}

\author{Min Zhang}
\email{z-m@tsinghua.edu.cn}
\affiliation{%
  \institution{Department of Computer Science and Technology, Institute for Internet Judiciary, Tsinghua University}
  \city{Beijing}
  \country{China}
}

\renewcommand{\shortauthors}{Su, et al.}

\begin{abstract}
Legal case retrieval aims to help legal workers find relevant cases related to their cases at hand, which is important for the guarantee of fairness and justice in legal judgments. While recent advances in neural retrieval methods have significantly improved the performance of open-domain retrieval tasks (e.g., Web search), their advantages haven’t been observed in legal case retrieval due to their thirst for annotated data. As annotating large-scale training data in legal domains is prohibitive due to the need for domain expertise, traditional search techniques based on lexical matching such as TF-IDF, BM25, and Query Likelihood are still prevalent in legal case retrieval systems. While previous studies have designed several pre-training methods for IR models in open-domain tasks, these methods are usually suboptimal in legal case retrieval because they cannot understand and capture the key knowledge and data structures in the legal corpus. To this end, we propose a novel pre-training framework named Caseformer that enables the pre-trained models to learn legal knowledge and domain-specific relevance-matching patterns in legal case retrieval without any human-labeled data. This framework is designed to support both dense retrieval models and neural re-ranking models. Through three unsupervised learning tasks, Caseformer is able to capture the special language, document structure, and relevance-matching patterns of legal case documents, making it a strong backbone for downstream legal case retrieval tasks. Experimental results show that our model has achieved state-of-the-art performance in both zero-shot and fine-tuning settings. Also, experiments on both Chinese and English legal datasets demonstrate that the effectiveness of Caseformer is language-independent in legal case retrieval.
\end{abstract}



\maketitle

\section{Introduction}
Legal case retrieval helps legal workers find relevant cases related to their cases at hand, which is important for the fairness and justice of legal judgments. As of today, most legal case retrieval systems still rely on simple bag-of-words retrieval models to retrieve documents based on users’ queries~\cite{rosa2021yes}. While recent advances in neural retrieval methods have significantly improved the performance of open-domain retrieval tasks (e.g., Web search)~\cite{karpukhin2020dense, nogueira2019passage}, their advantages haven’t been observed in legal case retrieval due to their thirst for annotated data. The training of state-of-the-art neural retrieval models usually requires millions or even billions of annotated query-document pairs to achieve desired effectiveness and reliability. In the legal domain, however, creating such large-scale training data is prohibitive due to the need for domain experts as assessors. For instance, in countries that adopt the civil law system \footnote{Civil law is the most widely adopted legal system in the world. It refers to structuring legal systems around broad codes and detailed statutes that determine individuals' rights and obligations.} (e.g., Germany, Japan, and China), prior cases are not required to be involved in judgment, which means that there are no similar cases recorded in judgment documents, extra efforts are needed to create annotated datasets for the training of neural retrieval models. In China, the largest case retrieval dataset LeCaRD ~\cite{ma2021lecard} only contains 107 labeled queries, which is far from enough to train an effective neural retriever. Therefore, statistical ranking models based on lexical matching such as TF-IDF~\cite{ramos2003using}, BM25~\cite{robertson2009probabilistic}, and Query Likelihood~\cite{zhai2008statistical} are still the mainstream techniques adopted by legal case retrieval systems~\cite{rosa2021yes,ma2021retrieving}.

To address the lack of supervision data in open-domain retrieval tasks such as Web search, pre-training methods that initialize neural retrieval models with unsupervised training signals have attracted much attention. As the pre-training and fine-tuning paradigms have achieved state-of-the-art performance in NLP tasks~\cite{devlin2018bert, vaswani2017attention, yang2019xlnet, liu2019roberta, yasunaga2022linkbert}, the IR community begins to explore pre-training methods tailored for IR~\cite{ma2021prop, ma2021pre, ma2021b, chang2020pre,fan2021pre,chen2022axiomatically, su2023thuir2, su2023wikiformer}. These PLMs are usually pre-trained on general domain corpus such as Wikipedia and have achieved better performance compared with their original versions such as BERT~\cite{devlin2018bert} and RoBERTa~\cite{liu2019roberta} in most retrieval and re-ranking tasks.

Unfortunately, existing PLMs tailored for general IR tasks do not fit the needs of legal case retrieval due to their incapability of legal document understanding. The definition of information relevance in the legal field is different from general ad-hoc retrieval tasks~\cite{ma2021lecard,shao2023understanding}. Besides the text similarity information used by open-domain retrieval models, case relevance in legal retrieval cares more about the legal similarity and relationships between legal elements~\cite{shao2023understanding}.  
\newadd{While some researchers have explored the possibility of adapting existing Pre-trained Language Models (PLMs) for legal data~\cite{chalkidis2020legal,xiao2021lawformer,li2023sailer}, a significant research gap exists in the modeling of legal relevance across different legal case documents. 
This aspect is particularly important for legal case retrieval because legal documents are interconnected by charges and cases with similar properties.
Given this, the current approaches, though advanced, fall short of fully harnessing the potential of PLMs for legal case retrieval tasks, highlighting a substantial opportunity for enhanced methodological development and improved performance in this specialized field.}

To this end, we propose a novel pre-train framework named Caseformer that enables the pre-trained models to learn legal knowledge and relevance-matching patterns between legal cases from raw legal corpora without any human-labeled data. This pre-training framework is designed to support both dense retrieval models and neural re-ranking models. Specifically, we propose three pre-training tasks: 1) Legal LAnguage Modeling (LAM), 2) Legal Judgment Prediction (LJP), and 3) Factual Description Matching (FDM). In the LAM task, we train the model to internalize the distinctive linguistic patterns and characteristics of the legal domain. In the LJP task, we train the model to measure relevance and connections between cases based on their similarity in legal judgments. 
Then, in the FDM task, we further train the model to measure case relevance based on the similarity between the fact descriptions in different case documents. 
Through these three pre-training tasks, Caseformer is able to capture domain-specific linguistic patterns, structures, and relevance-matching patterns across legal case documents, making it a strong backbone for downstream legal case retrieval tasks. Experimental results on three legal case retrieval datasets (both in English and Chinese) and two legal case relevance judgment datasets (both in English and Chinese) show that the re-ranking and retrieval models based on Caseformer can both achieve state-of-the-art performance in zero-shot and fine-tuning settings.

To summarize, the contributions of this paper are as follows:

\begin{itemize}

\item We propose a novel pre-training framework, Caseformer\footnote{We open source the entire project code in this link: https://github.com/oneal2000/Caseformer}, to solve the data-hungry problem of existing neural retrieval and re-ranking models in legal case retrieval scenarios. 
\item We propose three pre-training objectives that enable the proposed models to capture legal case documents' special language features, structure information, and relevance patterns between legal case documents.
\item We evaluate the performance of our framework on multiple legal case retrieval datasets, and the results show that Caseformer outperforms baselines in various settings.

\end{itemize}

\section{Related Work}
\subsection{Pre-training Methods for IR}

As pre-trained language models have achieved great success in the NLP field, the IR community begins to utilize PLMs to solve downstream IR tasks~\cite{li2023thuir,li2023towards,chen2023thuir,li2023constructing,ye2023relevance}, and design pre-training methods tailored for information retrieval~\cite{ma2021prop, ma2021pre, ma2021b, chang2020pre,fan2021pre,guo2022webformer,chen2022axiomatically,su2023caseformer,su2023thuir2,li2023sailer}.
The key idea of existing IR pre-training methods is to construct pseudo-relevant query-document pairs from unlabeled corpora. For example, \citeauthor{chang2020pre}~\cite{chang2020pre} designed three pre-training tasks based on Wikipedia: Inverse Cloze Task (ICT), Body First Selection (BFS), and Wiki Link Prediction (WLP). In these tasks, a sentence from a passage is randomly selected as a query, and the selected passage is defined as the corresponding relevant document. Also using Wikipedia as the pre-training corpus, ~\citeauthor{ma2021pre}~\cite{ma2021pre} utilizes the hyperlinks and their corresponding anchor text to train a re-ranking model named HARP. Webformer~\cite{guo2022webformer} utilizes the structural information of Wikipedia web pages and designs four pre-training tasks. Instead of using Wikipedia as the training corpus, PROP~\cite{ma2021prop} and B-PROP~\cite{ma2021b} are pre-trained on the plain text by the Representative Words Prediction (ROP) task. They assume that a sampled word set from a document with a higher query likelihood score is more “representative” of that document. Based on this assumption, they train the model to predict pairwise preference between two sampled word sets and achieve state-of-the-art performance.

{In summary, the pre-training objectives of current Information Retrieval Pre-trained Language Models (IR PLMs) are mostly designed for open domain tasks without special focus on any types of documents or domains. However, as shown in this paper, pre-training retrieval models without considering the special structures and characteristics of legal documents often lead to suboptimal performance in legal case retrieval. This motivates us to study how to construct and incorporate legal domain knowledge into the pre-training of legal retrieval models.}

\subsection{Legal Domain Pre-training}

As PLMs pre-trained in generic domains don’t work well on legal tasks, several studies have explored the possibility of constructing legal-specific pre-training models for legal tasks. ~\citeauthor{xiao2021lawformer}~\cite{xiao2021lawformer} proposed a legal domain pre-trained language model named Lawformer which is initialized by Longformer~\cite{beltagy2020longformer} as the basic encoder. Lawformer is pre-trained on millions of case documents published by China Judgments Online ~\footnote{https://wenshu.court.gov.cn} and has good performance after fine-tuning on downstream tasks. However, Lawformer is essentially a re-training of existing PLMs on the legal documents, which limits its capability in modeling domain-specific data structures.

\citeauthor{chalkidis2020legal}~\cite{chalkidis2020legal} propose a legal domain pre-trained model named Legal-BERT. They explore three strategies for using BERT to solve legal tasks: 1) use the original BERT directly, 2) adapt the original BERT by additional training on the legal domain, and 3) pre-train BERT from scratch on legal corpora. They found that further training in the legal domain is better than using the original BERT directly. Nonetheless, Legal-BERT directly uses the official BERT code in the pre-training stage and no changes have been made to adapt the unique characteristics of the legal field.

The most related study to this paper is the legal PLM proposed by ~\citeauthor{li2023sailer} ~\cite{li2023sailer} named SAILER.
\newadd{SAILER is designed to model a single legal case document through the encoder-decoder architecture. This architecture is adept at modeling and capturing the dependency between the Fact Description\footnote{The Fact Description section describes the facts and circumstances of a legal case.} section and other sections within a legal case document, which allows SAILER to leverage the logical relationships in the structures of a single legal document. 
However, it's crucial to note that SAILER's focus is predominantly on modeling the representation of individual legal case documents. It does not extend to examining the interrelationships between different legal documents. This limitation presents a gap in the current approach, as understanding the relevance among different legal documents is essential for comprehensive legal analysis and case retrieval. Our work aims to bridge this gap by proposing a method that not only accounts for the intricate details within individual case documents but also explores and models the relationships between different legal documents.

In summary, while existing work has successfully applied PLMs to the legal domain, they notably fall short in capturing the legal-level relevance among various legal case documents. This oversight underscores a significant opportunity for advanced methodological development and enhanced performance in the legal field. Our research is strategically positioned to bridge this gap. By employing three unsupervised tasks, we aim to enable PLMs to effectively model the relevance between legal documents. 
}

\begin{figure}[t]
\centering
    \includegraphics[width=0.8\textwidth]{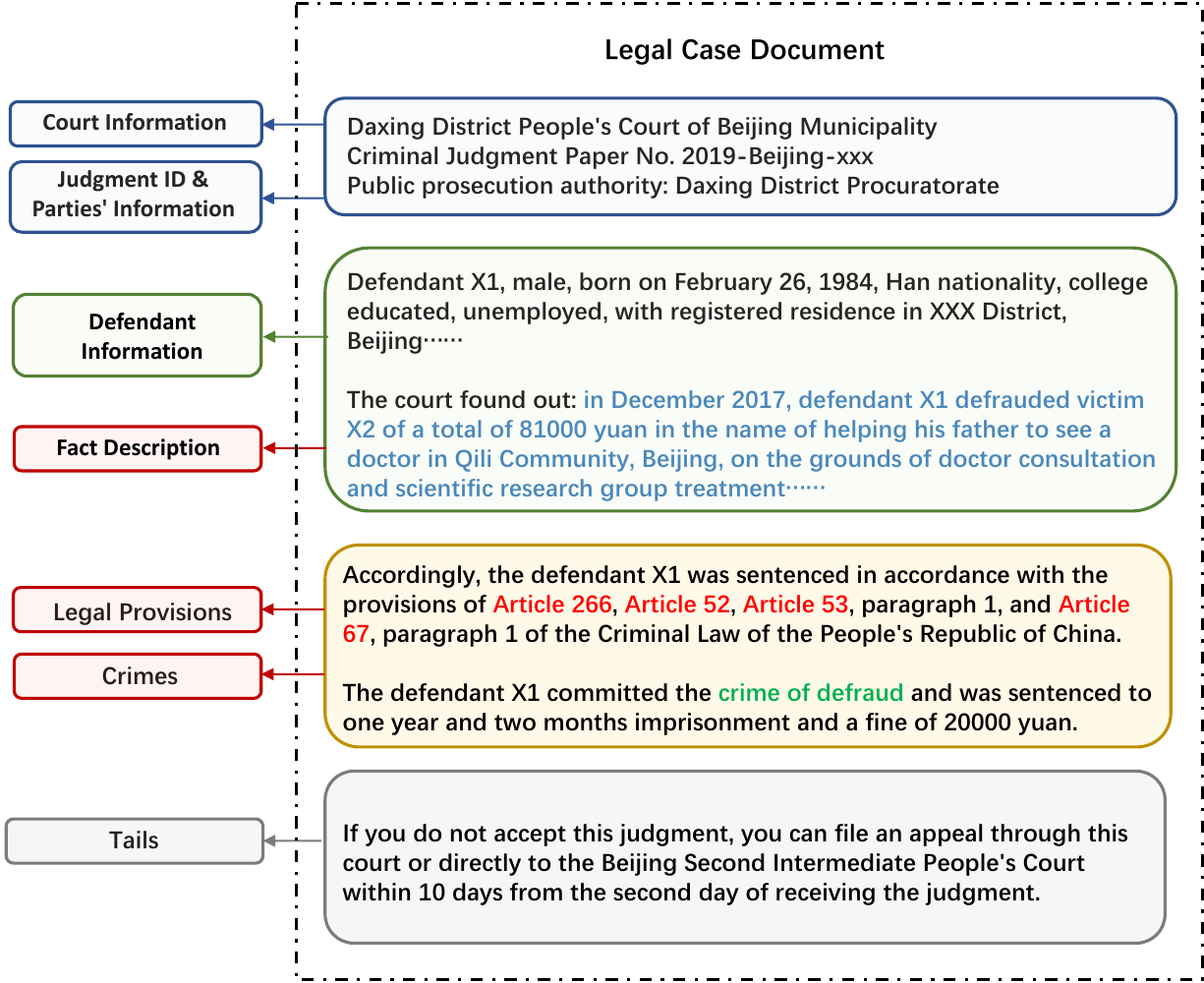}
    \caption{{An example of the writing organization of legal documents and their semi-structured information. The LJP and FDM tasks utilize three types of information: Fact Description, Legal Provisions, and Crimes which are highlighted in blue, red, and green fonts respectively.}}
    \label{structure}
\end{figure}
\section{Problem Formulation}

{The legal case retrieval task aims to retrieve relevant cases (represented by case documents) given the fact description of an unjudged query case. More specifically, given a query case $q$ and a set of candidate cases $C=\{c_1, c_2, ..., c_n\}$, where $n \in N^+$, let $r_i$ be the Bernoulli variable indicating whether $c_i$ is relevant to $q$, then the task of legal case retrieval is to retrieve a set of cases $S=\{c_j | r_j=1\}$.}

As shown in figure~\ref{structure}, a candidate legal case document usually consists of the following parts:

\begin{itemize}

\item \textbf{Court information} which provides detailed information about the court which produces the document. It typically includes the name of the court, the case number, the presiding judge's name, and the date of judgment.

\item \textbf{Defendant information} which provides information about the individual or entity against whom the legal action is being taken. It usually includes the defendant's full name, gender, date of birth, nationality, and other relevant identifying details.

\item \textbf{Fact Description} which describes the facts and circumstances of the case. It includes a comprehensive account of events, actions, or situations that led to the legal dispute. The fact description part of a well-written case document is usually clear, concise, and objective.

\item \textbf{Legal Provisions} which describes the relevant laws, statutes, regulations, or legal provisions that apply to the case. It may include references to specific sections or articles of the law that are relevant to the issues at hand. 

\item \textbf{Crimes} which describes the specific crimes or offenses that the defendant is accused of committing which serves to identify the charges against the defendant.

\end{itemize}

{In legal case retrieval practice, the query (q) and candidate documents typically comprises only the Fact Description part. Our work adopts this configuration, assuming that queries and candidate documents are facts descriptions extracted from legal case documents. In most cases, the length of the fact description section in a legal case is smaller than the maximum input length of the pre-trained model. For the rare cases where the fact description exceeds the maximum input length, we utilize a truncated approach to handle it (details can be found in Section ~\ref{section:implementation}).}

\begin{figure}[t!]
\centering
    \includegraphics[width=\textwidth]{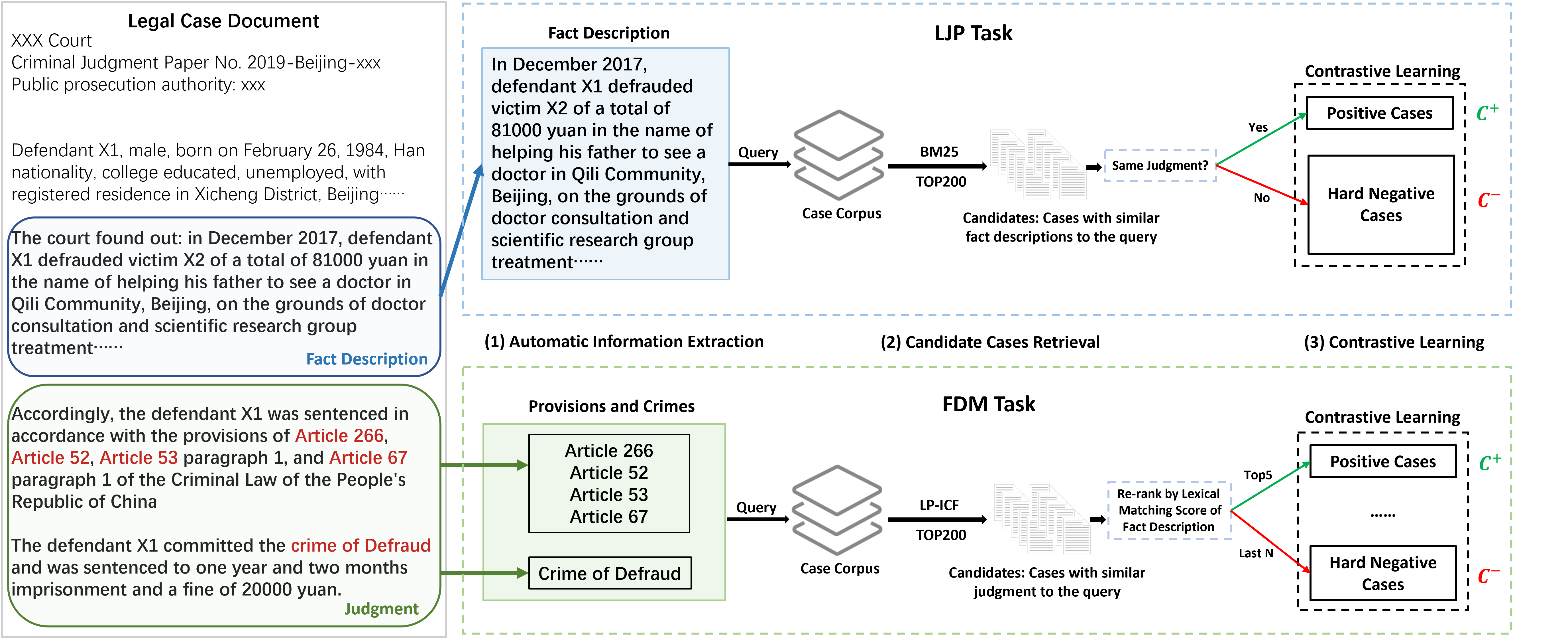}
    \caption{Illustration of the proposed pre-training tasks of Caseformer. Generally, there are three main stages: (1) Automatic Information Extraction, (2) Candidate Cases Retrieval, (3) Contrastive Learning.}
    \label{all_task}
\end{figure}

\section{Methodology}
In this section, we analyze the abilities of an ideal legal case retrieval model and discuss how we propose different pre-training tasks accordingly. To be specific, we introduce the model architecture in \S\ref{section:model} and the details of training process in \S~\ref{section:task1}, \S\ref{section:task2}, and \S\ref{section:task3}. The training process include three pre-training tasks: Legal LAnguage Modeling (LAM), Legal Judgment Prediction (LJP), and Factual Description Matching (FDM). And then in section \S\ref{section:final}, we introduce the overall training objective of our framework.

\qa{

\subsection{Model Architecture}
\label{section:model}
In practical retrieval systems, a two-stage pipeline is usually adopted to balance the overall effectiveness and efficiency. 
The first stage is the retrieval stage, where the relevant documents are recalled from an extensive corpus. 
This is followed by the re-ranking stage, where the documents recalled in the first stage are re-ranked according to their relevance to the query. 
The initial retrieval stage aims to swiftly find potentially relevant documents within the entire corpus.
The re-ranking stage, although more time-consuming, enables a more precise evaluation of each document's relevance to the query.

Existing Pre-trained Language Model (PLM) based search methods can be typically categorized into two architectures: dual-encoder and cross-encoder. Dual-encoders encode the query and candidate documents separately without considering the token-level interactions. As the corpus can be pre-encoded into dense representation vectors, dual-encoders are widely used in the first-stage retrieval. Cross-encoders, in contrast, concatenate queries and documents as a single input for the PLM, enabling detailed token-level interaction, thus enhancing the ranking accuracy. However, due to significant inference latency, cross-encoders are confined to re-ranking tasks within smaller datasets.

Applying this two-stage retrieval approach is also beneficial for legal case retrieval. The first stage ensures that the system can quickly process requests by filtering through a large legal case corpus. The second stage, while more time-consuming, enhances the quality of the results by carefully evaluating the relevance of each case. Therefore, we introduce both dual-encoder and cross-encoder architectures: Caseformer-Retriever and Caseformer-Re-ranker, designed for each stage respectively which are shown in figure~\ref{fig:dense} and figure~\ref{fig:cross}. This section will introduce the architectures of these two models.

\begin{figure}[t!]
\centering
    \includegraphics[width=0.99\textwidth]{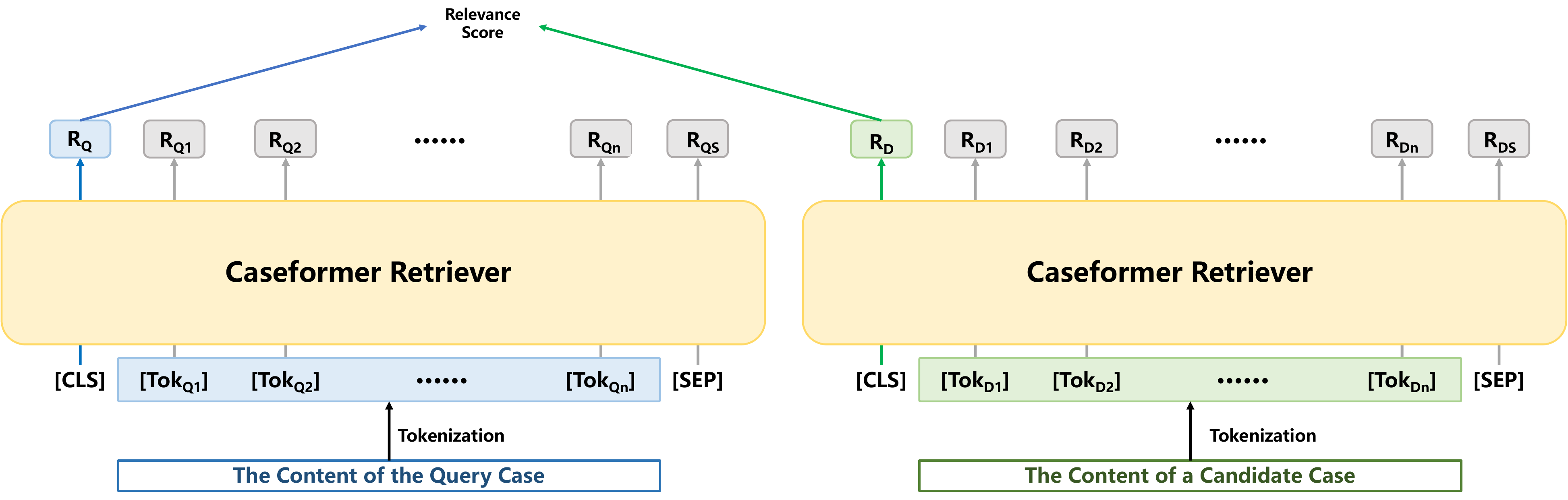}
    \caption{Illustration of the Caseformer-Retriever model's architecture. The process begins with word piece tokenization, appending special tokens [CLS] at the start and [SEP] at the end. Then the query case and the candidate case are encoded into dense vector representations. Subsequently, the relevance score is calculated based on these representations.}
\label{fig:dense}
\end{figure}

\subsubsection{Caseformer Retriever}
\label{section:dense_architecture}
We use the Transformer-encoder architecture (structurally the same as BERT~\cite{devlin2018bert}) for the implementation of Caseformer-Retriever.
Following the BERT's tokenization methodology~\cite{devlin2018bert}, we utilize word piece tokenization to convert the input text into discrete tokens. This tokenization process involves appending a [CLS] token at the start and a [SEP] token at the end of the token sequence. These tokens are then inputted into the Transformer-encoder~\cite{vaswani2017attention} to generate a contextualized embedding vector for each token. Following the setting of DPR~\cite{karpukhin2020dense}, the embedding corresponding to the [CLS] token serves as the comprehensive representation of the legal case. This encoding process of a legal case $C$ can be formulated as follows:

\begin{center}
\begin{equation}
\label{eq:tokenize}
{Input\_ids} = [{CLS}] \, {tokenizer}({C}) \, [{SEP}] 
\end{equation}
\end{center}

\begin{center}
\begin{equation}
\label{eq:rep}
 {Rep}(C) = {transformer}_{[CLS]}(Input\_ids) 
\end{equation}
\end{center}

\vspace{3mm}

\noindent where $tokenizer(x)$ utilizes word piece tokenization to convert the input text $x$ into discrete tokens, $transformer_{[CLS]}(\cdot)$ first encode the input with a transformer model and then extract the embedding vector of the [CLS] as the final representation of the input data. After acquiring the representations, we regard the inner product of two cases ($C_i$ and $C_j$) as their relevance score:

\begin{center}
\begin{equation}
\label{score_dense_1}
\begin{split}
s(C_i, C_j) = Rep(C_i)^\top \cdot Rep(C_j)
\end{split}
\end{equation}
\end{center}
\vspace{3mm}

\noindent In summary, the primary objective of the Caseformer-Retriever model is to model each legal case into a dense vector representation. The training process for the Caseformer-Retriever focuses on refining these legal case representations to optimize the retrieval effectiveness, which is detailed in \S~\ref{section:task1}, \S\ref{section:task2}, and \S\ref{section:task3}.

\subsubsection{Caseformer Re-ranker}

\begin{figure}[t!]
\centering
    \includegraphics[width=0.7\textwidth]{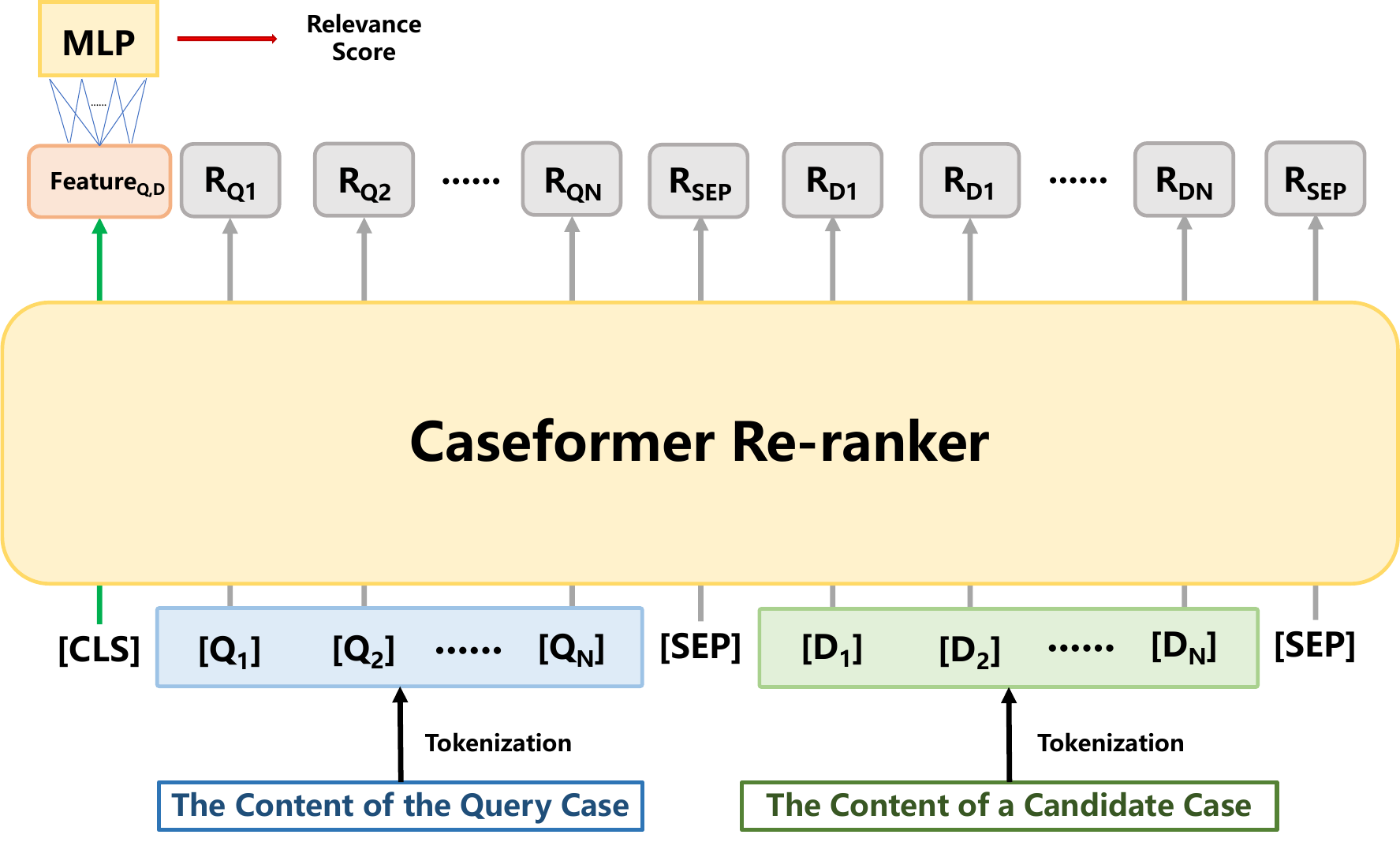}
    \caption{Illustraion of the Caseformer-reranker architecture. The process begins with word piece tokenization, appending special tokens [CLS] at the start, [SEP] for separation and at the end. The [CLS] token's embedding represents the interaction feature between the query and candidate case. A multi-layer perceptron (MLP) layer then maps this feature vector to a relevance score.}
\label{fig:cross}
\end{figure}

We use the Transformer-encoder architecture for the implementation of Caseformer-reranker. 
Following the BERT's tokenization methodology~\cite{devlin2018bert}, we utilize word piece tokenization. 
This tokenization process begins with adding a [CLS] token at the beginning, a [SEP] token to seperate the tokenized query and candidate case, and a [SEP] token at the end of the token sequence. 
These tokens are then inputted into the Transformer-encoder~\cite{vaswani2017attention} to generate a contextualized embedding vector for each token. 
Following the setting of~\cite{gao2021rethink,dai2019deeper}, the embedding corresponding to the [CLS] token serves as the representation of the token-level interaction feature between the query and the candidate case. 
Subsequently, a multi-layer perceptron (MLP) layer maps the relevance feature vector to its corresponding relevance score. 
This process can be formulated as follows:

\begin{equation}
\label{cross_input}
X_{C_Q,C_D} = [CLS] tokenizer(C_Q) [SEP] tokenizer(C_D) [SEP] 
\end{equation}

\begin{equation}
\label{score_cross}
f(C_Q, C_D) = transformer_{[CLS]}(X_{Q,D}) 
\end{equation}

\begin{equation}
\label{score_cross}
s(C_Q, C_D) = MLP(f(C_Q, C_D)) ) 
\end{equation}

\vspace{3mm}

\noindent where $C_Q$ is the query case and $C_D$ is the candidate case, $tokenizer(x)$ utilizes word piece tokenization to convert the input text $x$ into discrete tokens, $MLP(\cdot)$ is a multi-layer perceptron that projects the the relevance feature vector $f$ to a relevance score $s$. The training process for the Caseformer-Reranker is detailed in \S~\ref{section:task1}, \S\ref{section:task2}, and \S\ref{section:task3}.

}

\subsection{Legal LAnguage Modeling (LAM) Task: Understanding Legal Language}
\label{section:task1}

{As discussed previously, an ideal legal case retrieval model should have the ability to capture and understand the domain language used in legal documents. As illustrated in Figure~\ref{structure}, legal documents often contain specialized terminology, expressions, and content structures that are rarely observed in general domain documents. We consider these specific professional terms, expressions, and writing structures within the legal field as legal language.}

{To address the limitations of existing PLMs in legal case retrieval, we propose a pre-training task named Legal LAnguage Modeling (LAM). Specifically, we first pre-train the model on official law books (e.g., official criminal code, official judicial interpretation, etc.) with the Mask Language Modeling (MLM) task. For each document in the corpus, we tokenize the document and then divide the tokenized document into input sequences $X = [x_1, x_2,x_3, …, x_n]$, where $n$ is the maximum input length of the model. We then use a dynamic masking strategy~\cite{liu2019roberta} to randomly replace the tokens in $X$ with the special token [MASK]. The MLM loss L can be calculated as:}

\begin{equation}
\label{mlm_loss}
\mathcal{L}_{MLM} = -\sum_{\hat{x} \in m(X)} \log(P(\hat{x} | X_{\backslash m(X)}))
\end{equation}
\vspace{3mm}

{\noindent where $X$ denotes the input sequence, $m(X)$ and $X_{\backslash m(X)}$  are the masked word set and the rest words in $X$, respectively, $ \log(P(\hat{x} | X_{\backslash m(X)}))$ is the model predicted probability of token $\hat{x}$. The model is trained to minimize this MLM loss by updating its parameters through backpropagation and gradient descent optimization algorithms. }The official law books we used here are issued by the government of a country with strict language organization and reliable content. They usually contain a comprehensive range of professional terms and expressions utilized in the legal domain. Therefore, through conducting MLM training on this corpus, we want the model to acquire an accurate understanding of the meaning associated with professional terms and expressions used in the legal field.

{To enhance the model's acquisition of legal knowledge and its ability to adapt to the distinctive writing structure within the legal domain, we additionally conduct pre-training on legal case documents with the MLM task. These legal documents encompass a wealth of legal knowledge, such as the exposition of fundamental facts, legal provisions, and criminal offenses. By training the model on these documents, we want to adapt it to the specific writing structure characteristic of the legal field, while also acquiring a comprehensive understanding of legal knowledge.}

Overall, the idea of the LAM task is to pre-train retrieval models with MLM tasks on legal documents that provide comprehensive definitions and explanations of legal terminology and concepts. With LAM, our goal is to enhance the model's ability in understanding legal language in fine grains so that it can better serve downstream training and applications.

\begin{figure}[t!]
\centering
    \includegraphics[width=\textwidth]{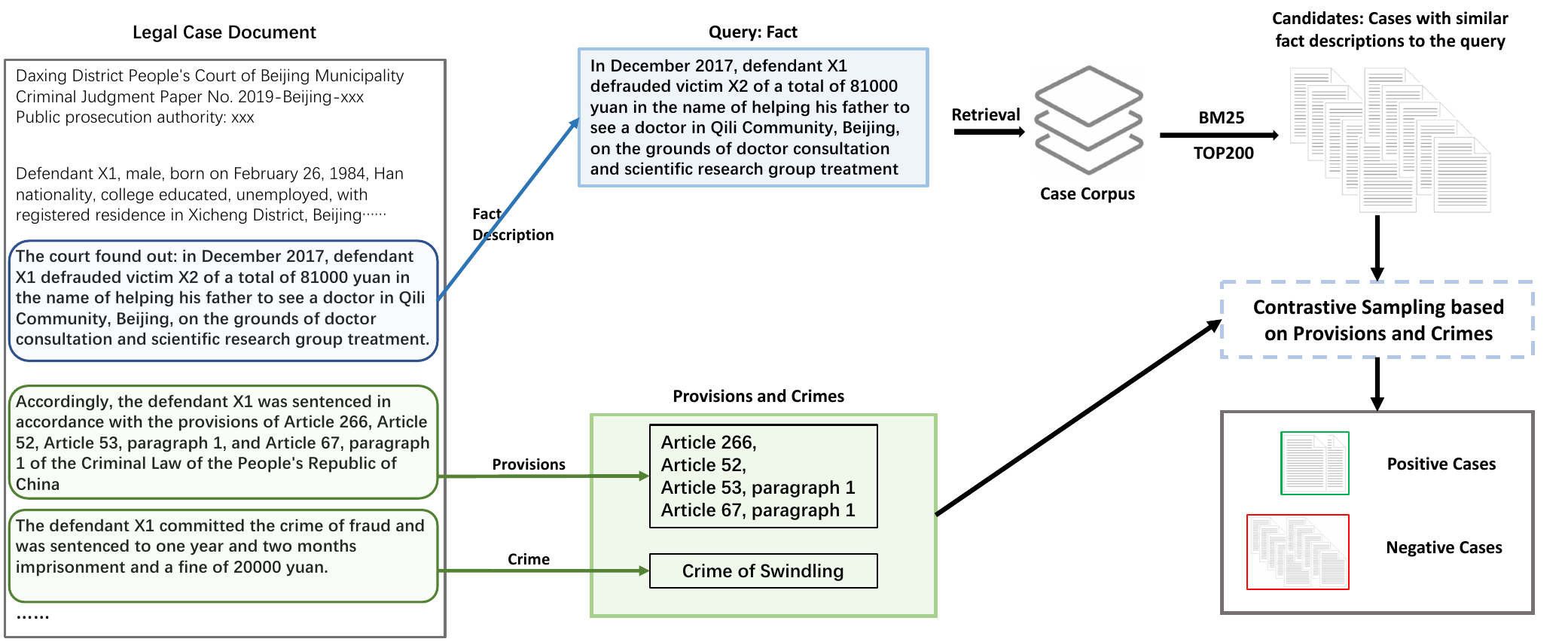}
    \caption{The contrastive sampling strategy of the LJP task.}
    \label{pic:ljp}
\end{figure}

\subsection{Legal Judgment Prediction (LJP) Task: Measuring Judgment Similarity}
\label{section:task2}

Retrieval models pre-trained on general domain data are usually good at matching documents according to their semantic similarity in text. However, in the context of legal case retrieval, the models should not only consider semantic similarity but also evaluate the legal-level similarity between cases. This distinction arises from the unique needs of the legal field, where the assessment of case relevance consider both semantic and legal aspects. For example, consider the fact descriptions of the following cases,

\vspace{4mm}
\vspace{2mm}

\noindent \textbf{Case 1}:\\ \textit{On March 10, 2022, on X Street in K City, \textbf{two} people with baseball hats gathered to fight after drinking. The police quickly arrived at the scene, controlled the people involved in the fight, and avoided further escalation of the situation. This case finally resulted in the \textbf{reconciliation} of those two people.} 

\vspace{2mm}
\vspace{2mm}

\noindent \textbf{Case 2}: \\\textit{On March 10, 2022, on X Street in K City, \textbf{twenty} people with iron baseball bats gathered to fight after drinking. The police quickly arrived at the scene, controlled the people involved in the fight, and avoided further escalation of the situation. This case finally resulted in \textbf{two deaths}}.

\vspace{2mm}
\vspace{4mm}

\noindent In the first case, the situation was quickly resolved, resulting in reconciliation between the parties involved. This suggests that it may not be a criminal case and might have been handled through alternative means such as mediation or civil proceedings. On the other hand, the second case, while having a similar fact description in terms of time, location, and gathering of individuals, it finally escalated into a major criminal case with two deaths involved. It can be seen that some cases with similar semantics could be fundamentally different at the legal level. 

To teach the model to understand case relevance from legal perspectives, we train the model on legal case documents by proposing an unsupervised contrastive learning task named Legal Judgment Prediction (LJP). As shown in Figure ~\ref{structure}, a standard case document consists of several parts including factual descriptions, legal provisions, and the crimes of the case which can be automatically extracted based on the writing structure of each case. The proposed Legal Judgment Prediction (LJP) task is illustrated in Figure ~\ref{all_task}. The basic assumption of LJP is that, \textbf{among cases with similar fact descriptions, cases with the same judgments are usually relevant to each other}. Specifically, we train the model to select cases with the same crimes and provisions from a series of cases with similar factual descriptions. Firstly, given a corpus consisting of legal case documents, we extract the factual description, committed crimes, and legal provisions of each case. Then a case $Q$ is randomly selected from a case collection corpus and its factual description is used as the query. Based on the query $Q$, the BM25 method is adopted to compute the similarity between the fact descriptions of the query and the candidates, and recall the top 200 similar cases in terms of fact description. The recalled cases set is defined as $C$. \textbf{For each case $c_i$ in $C$, if the crimes and legal provisions of $c_i$ are the same as the query case $Q$, then we treat $c_i$ as a positive example and add it to the set $C^+$, if not, $c_i$ is defined as a negative example and added to the set $C^-$.}

After sampling the positive and negative examples, we use a contrastive learning strategy to train both the re-ranking model and the retrieval model. \textbf{For the re-ranking model}, we use the cross-encoder architecture~\cite{nogueira2019passage} to compute the relevance score between two legal case documents $c_i$ and $c_j$:

\begin{equation}
\label{cross_input}
X_{ij} = [CLS] c_i^f [SEP] c_j^f [SEP] 
\end{equation}

\begin{equation}
\label{score_cross}
s(c_i, c_j) = MLP(transformer_{[CLS]}(X_{ij}) ) 
\end{equation}

\noindent where $c_i^f$ is the factual description extracted from the $c_i$, $transformer_{[CLS]}(\cdot)$ first encode the input with a transformer model and then output the embedding vector of the [CLS] as the final representation of the input data. $MLP(\cdot)$ is a multi-layer perceptron that projects the [CLS] embedding to a relevance score $s$.

\textbf{For the retrieval model}, we use the dual-encoder architecture~\cite{karpukhin2020dense} to compute the dot product between two embedding vectors as the relevance score:

\begin{equation}
\label{cross_input_dual}
X(c) = [CLS] c^f [SEP]
\end{equation}

\begin{equation}
\label{embedding}
Emb(X) =  transformer_{[CLS]}(X) 
\end{equation}

\begin{center}
\begin{equation}
\label{score_dense}
\begin{split}
s(c_i, c_j) = Emb(X(c_i))^\top \cdot Emb(X(c_j))
\end{split}
\end{equation}
\end{center}

\noindent where $c^f$ is the factual description extracted from the input case $c$, $transformer_{[CLS]}(\cdot)$ outputs a contextualized vector for each token and we select the "[CLS]" vector as the embedding vector of a case. In Equation ~\ref{score_dense}, we regard the inner products of case embeddings as the relevance score $s$.

For the loss function, we use the Softmax Cross Entropy Loss~\cite{cao2007learning,ai2018learning,gao2021rethink} to optimize the re-ranking and retrieval model, which is defined as:
\begin{equation}
    \label{eq:LRP}
\begin{aligned}  
   & \mathcal{L}_{LJP}(Q,c^+,N) \\& = -\log_{}{    \frac{exp(s(Q,c^+))}{exp(s(Q,c^+) + \sum_{c^-\in N} exp(s(Q,c^-))}} 
\end{aligned}
\end{equation}

\noindent where $s$ is the relevance score function which is defined in Equation~\ref{score_cross} and Equation~\ref{score_dense} for re-ranking and retrieval models respectively. $Q$ is the query case, $c^+$ is a selected positive case and $N$ is the set of selected negative cases.

{Note that the matching of crimes and legal provisions may be influenced by typographical errors and non-standard writings, but in our dataset, such occurrences account for less than 1\%. We have also employed a range of methods to address this issue, and the specific implementation details can be found in our open-source code\footnote{https://github.com/caseformer/caseformer}. This problem is primarily an engineering problem and is not the focus of this paper.}

\subsection{Factual Description Matching (FDM) Task: Measuring Factual Similarity}
\label{section:task3}

\begin{figure}[h]
\centering
    \includegraphics[width=\textwidth]{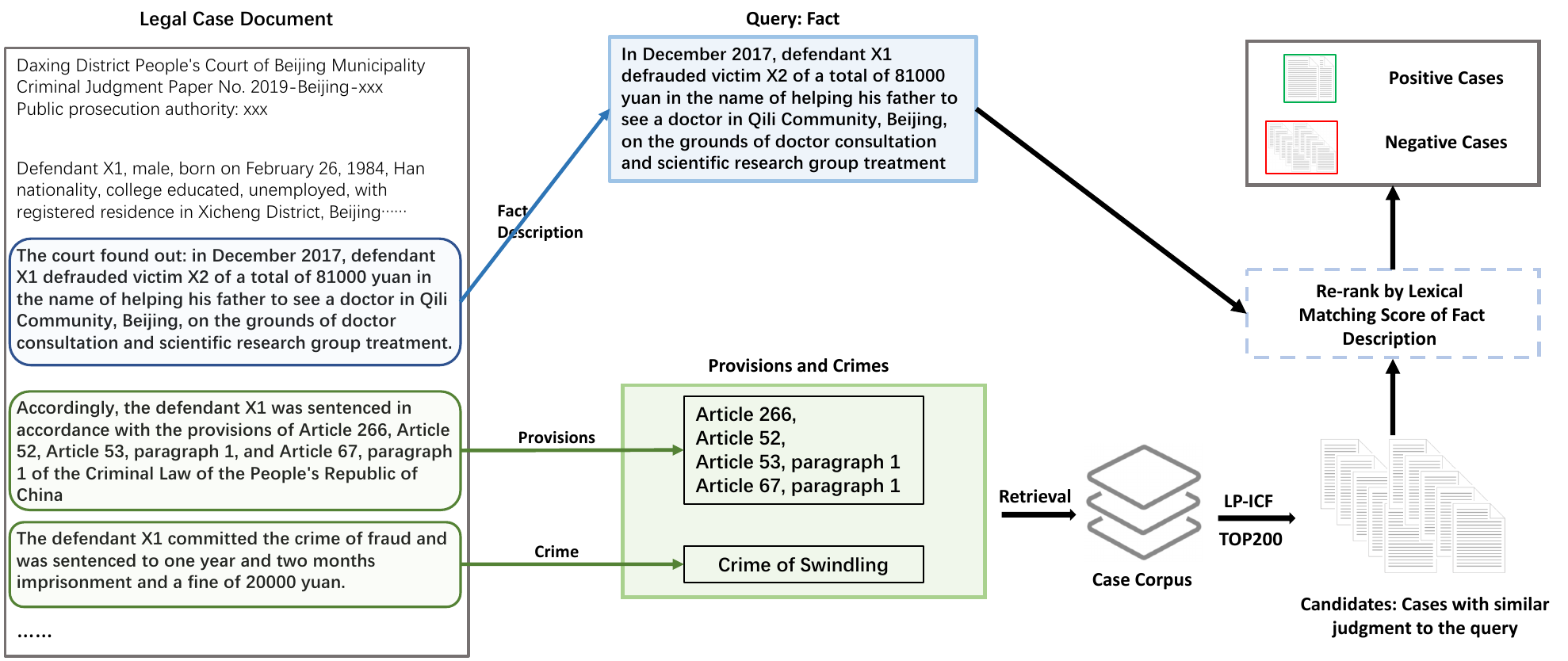}
    \caption{The contrastive sampling strategy of the FDM task.}
    \label{pic:fdm}
\end{figure}


With the LJP task, we train the model to better distinguish cases based on their judgment information. On the other hand, an ideal case retrieval model should also be able to distinguish relevant and irrelevant case documents based on factual description information of legal cases. Therefore, we propose the Factual Description Matching (FDM) task which is illustrated in Figure ~\ref{all_task}. Our assumption is that \textbf{in cases with similar judgments, cases with similar fact descriptions should be more relevant to each other than those that don't.} To be specific, we train the model to select relevant cases based on the fact description from a set of candidate cases with similar legal judgments. 
First, we propose a retrieval method named Legal Provision-Inverse Case Frequency (LP-ICF) to find cases with similar judgments. Specifically, given a legal case document, we extract its crimes (charges) and legal provisions from the judgment. Based on the crimes and legal provisions, the Legal Provision-Inverse Case Frequency (LP-ICF) method is defined as follows. Given a legal case document collection D, a case as the query ($c_i$), and a candidate case ($c_j$):

\begin{equation}
    \label{ICF}
 LP-ICF(c_i, c_j) = crime_{i,j} * \sum_{p\in P} log_{}{    \frac{|D|}{freq(p,D)}} 
\end{equation}

\noindent where $P$ is the set of overlapping provisions between $c_i$ and $c_j$, $|D|$ is the size of the collection $D$, $freq(p,D)$ is the number of the appearance of provision $p$ in the collection $D$, and $crime_{i,j}$ is set to 0 if there's no overlap in the crimes between $case_i$ and $case_j$ and otherwise, is set to 1. The LP-ICF method can be understood as follows: first, calculate the inverse case frequency (ICF) of each legal provision based on its appearance. Then, according to the overlap of the two cases in crimes and legal provisions, their similarity score is calculated by Equation~\ref{ICF}. In short, LP-ICF can recall a series of cases that are similar in judgment within a short time. 

Based on the LP-ICF method, given a case $Q$, we first recall the top 200 relevant cases that are similar in judgment by LP-ICF. The recalled cases set is defined as $C$. \textbf{For each case $c_i$ in $C$, we calculate the lexical relevance between the factual description of $c_i$ and $Q$ by BM25. Then randomly select a case from the top 5 of the BM25 ranking list as the positive example $c^+$ and select the last $\lambda$ cases of the BM25 ranking list (where $\lambda$ is an adjustable hyperparameter) and add them to the negative examples set $C^-$.}

After sampling the positive and negative examples, we use a contrastive learning strategy to train both the re-ranking model and the retrieval model. For the loss function, we still use the Softmax Cross Entropy Loss to optimize our model, which is defined as:
\begin{equation}
    \label{FRP_loss}
\begin{aligned}  
   & \mathcal{L}_{FDM}(Q,c^+,N) \\& = -\log_{}{    \frac{exp(s(Q,c^+))}{exp(s(Q,c^+) + \sum_{c^-\in N} exp(s(Q,c^-))}} 
\end{aligned}
\end{equation}

\noindent where $s$ is the relevance score function which is defined in Equation~\ref{score_cross} and Equation~\ref{score_dense} for re-ranking and retrieval models. $Q$ is the query cases, $c^+$ is the selected positive case and $N$ is the set of selected negative cases.

\subsection{Final Training Objective}
\label{section:final}

We combined the above three tasks as the final training objective, which is defined as follows:

\begin{equation}
\label{FRP_loss}
\begin{aligned}  
   & \mathcal{L}_{final}(Q,P_{LJP}, N_{LJP}, P_{FDM}, N_{FDM}) \\&= \mathcal{L}_{LAM}(Q) + \mathcal{L}_{LJP}(Q,P_{LJP},N_{LJP}) + \mathcal{L}_{FDM}(Q,P_{FDM},N_{FDM})
\end{aligned}
\end{equation}

\noindent where $Q$ is a case from the corpus, $P_{LJP}$ is the positive example generated from the LJP task, $N_{LJP}$ is the set of negative examples generated from the LJP task, and  $P_{FDM}$, $N_{FDM}$ are the positive example and negative examples generated from the FDM task. We conducted ablation experiments on all combinations of loss in Section~\ref{section:ablation}.

\section{Experiments}

\subsection{Experiment Setup}

\subsubsection{Pre-training Corpus}
\label{sec:pre-train-corpus}
We select two types of legal corpus, one in Chinese and one in English, as the pre-training corpora for our experiments. For the Chinese version of Caseformer under the Chinese criminal law system, we first pre-train the model on the official law books\footnote{All corpus are uploaded to our anonymous Github link: https://github.com/caseformer/caseformer} in the LAM task. Then in the LJP and FDM tasks, we pre-train the model on 5M case documents released by the Supreme Court of China. Based on this corpus, we generate around 800M pseudo-query-case pairs. For the English version of Caseformer, we first pre-train the model on the Indian Penal Code in the LAM task. Then in the LJP and FDM tasks, we pre-train the model on the ILSI~\cite{paul2022lesicin} dataset which contains 66,090 legal cases from several major Indian Courts. The fact description and legal provisions of each case are provided in ILSI. Based on this corpus, we generate around 13M pseudo-query-case pairs.

\subsubsection{Legal case retrieval Datasets}
We evaluate the performance of Caseformer on the following datasets. 

\begin{itemize}

    \item[-] \textbf{LeCaRD}~\cite{ma2021lecard} is the largest Chinese case retrieval dataset, consisting of 107 query cases and over 43000 candidate cases\footnote{Note that LeCaRD provides two types of re-ranking tasks: the number of candidate documents is 30 and 100 respectively. The original paper~\cite{ma2021lecard} and Lawformer~\cite{xiao2021lawformer} use the setting of 30. We choose the setting of 100 (following the setting of SAILER~\cite{li2023sailer}) which is more likely to distinguish the differences in models' performance.}. All the cases are adopted from criminal cases published by the Supreme People’s Court of China. {The queries and candidate documents in the LeCaRD dataset are the factual description part (introduced in Section 3) of a legal case.}

    \item[-] \textbf{CAIL-SCM}~\cite{xiao2019cail2019} is a case relevance judgment dataset provided by CAIL 2019. All the cases are published by China Judgments Online\footnote{https://wenshu.court.gov.cn/}, an official website of the Chinese Legal System. Each data is composed of one query case and two candidate cases. For each legal case document, the title and fact description is provided. {Both the queries and candidate documents in the CAIL-SCM dataset are the factual descriptions part of a legal case.}

    \item[-] \textbf{CAIL-LCR}~\footnote{https://github.com/china-ai-law-challenge/CAIL2022/tree/main/lajs} is a case retrieval dataset provided by CAIL 2022 consisting of 130 query cases and 100 candidate cases for each query case. {The queries and candidate documents in the LeCaRD dataset are the factual description part of a legal case.}

    \item[-] \textbf{COLIEE 2020 Task1} ~\cite{rabelo2021coliee} is an English version case retrieval task provided by COLIEE\footnote{https://sites.ualberta.ca/~rabelo/COLIEE2020/}. The training set contains 520 query cases and 200 candidate cases for each query case. The test set contains 130 query cases and 200 candidate cases for each query case. {Both the queries and candidate documents in the COLIEE 2020 dataset are complete legal case documents with all parts listed in Figure 1.}

\end{itemize}

\subsubsection{Baselines}\label{sec:baselines}

We consider four types of baselines for comparison, including traditional IR methods, pre-trained Language models on general domain data, PLMs tailored for IR, and pre-trained language models built with legal documents. 

\begin{itemize}
    \item \textbf{Traditional IR Methods}
    
    \begin{itemize}

        \item \textbf{QL}~\cite{zhai2008statistical} is a language model based on Dirichlet smoothing and has good performance on retrieval tasks.
        
        \item \textbf{BM25}~\cite{robertson2009probabilistic} is a highly effective retrieval model based on lexical matching that achieves good performance in retrieval tasks.
    \end{itemize}  

    For the implementation, we use the pyserini toolkit\footnote{https://github.com/castorini/pyserini}. For the hyperparameter of BM25, we set $k1 = 3.8$ and $b = 0.87$\footnote{This is the best hyperparameter we got after parameter searching.}. Note that in our experiments, we use the scores of the BM25 and QL models to re-rank the candidate documents, rather than re-ranking the whole corpus.
    
    \item \textbf{Pre-trained Models tailored for IR}

    \begin{itemize}

        \item \textbf{PROP}~\cite{ma2021prop} is a pre-trained model with cross-encoder architecture tailored for IR re-ranking tasks. It adopts the Representative Words Prediction (ROP) task to predict the pairwise preference between word sets. \footnote{As PROP and B-PROP have almost the same performance, we choose one of these two models as the baseline.} 
        
        \item \textbf{SEED}~\cite{lu2021less} is a pre-trained text encoder for dense retrieval that achieves state-of-the-art performance.

        \item \textbf{Condenser}~\cite{gao2021condenser}. Condenser ~\cite{gao2021condenser} is a state-of-the-art pre-training architecture for dense retrieval. It leverages skip connections to consolidate textual information into dense vectors.

        \item \textbf{coCondenser}~\cite{gao2021unsupervised}. CoCondenser is an enhanced version of Condenser that adds an unsupervised corpus-level contrastive loss to warm up the passage embedding space. 


    \end{itemize}
    
    As PROP, SEED, Condenser, and coCondenser have no available Chinese versions, we reproduce their work on the Chinese corpus described in section~\ref{sec:pre-train-corpus} based on their open-source training code and follow all settings provided in their paper~\cite{ma2021prop,lu2021less,gao2021condenser,gao2021unsupervised}.
    
    \item \textbf{General Domain Pre-trained Models}
    
    \begin{itemize}
        \item \textbf{BERT}~\cite{devlin2018bert} is a bi-directional Transformer based encoder that has a powerful ability on contextual text representations and achieves state-of-the-art performance on many NLP downstream tasks as well as IR tasks. 
        
        \item \textbf{RoBERTa}~\cite{liu2019roberta} shares the same architecture with BERT and is trained on a larger corpus through the MLM task. 
        
        \item \textbf{Chinese-BERT-WWM}~\cite{cui2021pre} is a BERT-based model pre-trained with Whole Word Masking (WWM) strategy in Chinese corpora. 

        \item \textbf{Chinese-RoBERTa-WWM}~\cite{cui2021pre} is a RoBERTa-based model pre-trained with Whole Word Masking (WWM) strategy in Chinese corpora. 

        \item {\textbf{text-embedding-ada-002}\footnote{https://platform.openai.com/docs/guides/embeddings}, an embedding model developed by OpenAI, serves as a powerful tool for text search, text similarity, and code search. It achieves SOTA performance across various datasets such as BEIR\cite{thakur2021beir}, SentEval\cite{conneau2018senteval}, etc.
        }

    \end{itemize}
    
    For the implementation of BERT, we use the Pytorch version BERT-base released by Google\footnote{https://github.com/google-research/bert}. For the implementation of RoBERTa, Chinese-BERT-WWM and Chinese-RoBERTa-WWM, we directly use their models released on Huggingface\footnote{https://huggingface.co/roberta-base, https://huggingface.co/hfl/chinese-bert-wwm, https://huggingface.co/hfl/chinese-roberta-wwm-ext}.

    \item \textbf{Legal Domain Pre-trained Models}
    
    \begin{itemize}
        \item \textbf{Legal-BERT}~\cite{chalkidis2020legal} is a BERT model pre-trained in the legal domain that directly uses the official BERT code in the pre-training stage.
        
        \item \textbf{BERT-XS}~\footnote{http://zoo.thunlp.org} is a legal domain BERT model trained on the Chinese criminal document corpus. 

        \item \textbf{Lawformer}~\cite{xiao2021lawformer} apply Longformer\cite{beltagy2020longformer} to initialize and train with the MLM task on the legal domain.

        \item \textbf{SAILER}~\cite{li2023sailer}. SAILER is a structure-aware pre-trained model for legal case representation. It utilizes the logical connections within a legal document's structure.

    \end{itemize}

\end{itemize}


\begin{table}[]
\caption{A comparative overview of our selected baseline models. This table categorizes various pre-trained language models based on their type (Retriever or Re-ranker), language support (English, Chinese, or Multilingual), presence of an MLP layer (indicated by \ding{51} for presence and \ding{55} for absence), and the primary purpose of the MLP layer.}
\label{tab:baselines}
{\footnotesize
\begin{tabular}{cccccc}

\toprule
                             & Model Name             & Language & Model Type             & MLP Layer  & MLP Purpose        \\
\toprule
\multirow{4}{*}{IR PLM}      & PROP                   & En \& Zh    & Re-ranker              & \ding{51}         & Semantic Relevance \\
                             & SEED                   & En \& Zh    & Retriever              & \ding{55}         &           -         \\
                             & Condenser              & En \& Zh    & Retriever              & \ding{55}         &            -        \\
                             & coCondenser            & En \& Zh    & Retriever              & \ding{55}         &             -       \\
\midrule
\multirow{5}{*}{General PLM} & BERT                   & En               & Retriever / Re-ranker & \ding{51}         & Next Sentence Prediction                \\
                             & RoBERTa                & En               & Retriever              & \ding{55}         &   -                 \\
                             & Chinese-BERT-WWM       & Zh               & Retriever / Re-ranker & \ding{51}         & Next Sentence Prediction                \\
                             & Chinese-RoBERTa-WWM    & Zh               & Retriever              & \ding{55}         &   -                 \\
                             & text-embedding-ada-002 & Multi         & Retriever              & \ding{55}         &   -                 \\
\midrule
\multirow{4}{*}{Legal PLM}   & Legal-BERT             & En               & Retriever / Re-ranker & \ding{51}         & Next Sentence Prediction                \\
                             & BERT-XS                & Zh               & Retriever / Re-ranker & \ding{51}         & Next Sentence Prediction                \\
                             & Lawformer              & Zh               & Retriever              & \ding{55}         &    -                \\
                             & SAILER                 & En \& Zh    & Retriever              & \ding{55}         &        -            \\
\midrule
\multirow{2}{*}{Ours}        & Caseformer Retriever   & En \& Zh    & Retriever              & \ding{55}         &        -            \\
                             & Caseformer Re-ranker   & En \& Zh    & Re-ranker              & \ding{51}         & Legal Relevance   \\
\toprule
\end{tabular}
}
\end{table}

It's beneficial to understand the characteristics and functionalities of our selected baselines. These include the model type (Retriever or Re-ranker), the range of language support (English, Chinese, or Multilingual), the incorporation of an MLP layer, and the specific purposes of these MLP layers. To provide a clear and comprehensive view of these attributes, a summarized comparison of these models is detailed in Table ~\ref{tab:baselines}.

\subsubsection{Implementation Details}
\label{section:implementation}


Our implementation details of Caseformer and other baselines are described as follows.

    \noindent \textbf{Caseformer and Baselines Implementations.} For the Chinese version of Caseformer, we initialize our re-ranking model with the Chinese-BERT-WWM\footnote{https://huggingface.co/hfl/chinese-bert-wwm} and retriever with Chinese-RoBERTa-WWM\footnote{https://huggingface.co/hfl/chinese-roberta-wwm-ext}. For the Engish version of Caseformer, we initialize our re-ranking model and retrieval model respectively with the BERT-base-uncased\footnote{https://huggingface.co/bert-base-uncased} and RoBERTa\footnote{https://huggingface.co/roberta-base} checkpoints from Huggingface. We set the hyperparameter $\lambda$ in the FDM task to $16$. For the implementation of Legal-BERT, BERT-XS, and Lawformer we use the checkpoints released by the original paper. As PROP, SEED, Condenser, and coCondenser have no available Chinese versions, we reproduce their work on Chinese corpora described in section~\ref{sec:pre-train-corpus} based on their open-source training code and follow all settings provided in their paper~\cite{ma2021prop,lu2021less,gao2021condenser,gao2021unsupervised}. For the BM25 and Querylikehood method, we use the pyserini toolkit\footnote{https://github.com/castorini/pyserini} with default hyperparameters.

    \noindent \textbf{Pre-training Settings.} In the LAM task, we follow the masking strategy of BERT. In the FDM and LJP tasks, we select every case from the corpus as the query case and generate positive cases and negative cases via the contrastive sampling strategy introduced in Section 4. We use the AdamW optimizer with a learning rate of $5e^{-6}$ and a warm-up ratio of $0.1$. In the LAM task, we set the masking ratio as $0.15$. In the LJP and FDM task, we set the maximum length of the query and document to 510 and truncate the rest. For the baseline PLMs, we follow all the pre-training settings in the original paper. For computing costs, we pre-trained our model on 8 Nvidia GeForce RTX 3090 GPUs for 120 hours.

\subsubsection{Statistical Significance Evaluation} For the significance test, we adopt Fisher’s randomization test ~\cite{fisher1936design,cohen1995empirical,box1978statistics} which is recommended for IR evaluation by previous work ~\cite{smucker2007comparison}.

\subsection{Experimental Results}
\subsubsection{\textbf{Caseformer Retriever}}
\begin{table}[t!]

\caption{The experimental results of the Caseformer retriever (dual-encoder architecture) and other baselines on LeCaRD and CAIL-LCR in the zero-shot setting. The best results are in bold. “*” denotes the result is significantly worse than Caseformer with $ p < 0.01 $ level. R@k indicates the Recall@k metric in this table.}
\label{dense_zero}
  
\centering

\begin{tabular}{lccc|ccc}
\toprule
& \multicolumn{6}{c}{\textbf{Zero-shot}} \\
\cmidrule(r){2-7}
                    & \multicolumn{3}{c}{\textbf{LeCaRD}}                 & \multicolumn{3}{c}{\textbf{CAIL-LCR}}                                   \\
                    \cmidrule(r){2-4} \cmidrule(r){5-7}
                    & {R@100}  & {R@200}  & {R@500}  & {R@100}  & {R@200}  & {R@500}  \\
                    \toprule
\textbf{BM25}       & \textbf{0.5154}          & \textbf{0.6781}          & 0.8249          & 0.4484*          & 0.6029*          & 0.7924          \\
\midrule
\textbf{BERT}       & 0.1538*          & 0.2216*          & 0.3351*          & 0.1784*          & 0.2475*          & 0.3448*          \\
\textbf{RoBERTa}    & 0.4912          & 0.6011*          & 0.7537*          & 0.5619          & 0.6889*          & 0.8092*          \\
\textbf{SEED}       & 0.3311*          & 0.4389*          & 0.6378*          & 0.4534*          & 0.5721*          & 0.7271*          \\
\textbf{Condenser}       & 0.3728*          & 0.5227*          & 0.6944*          & 0.4523*          & 0.5694*          & 0.7243*          \\
\textbf{coCondenser}       & 0.4098*          & 0.5515*          & 0.7240*          & 0.4505*          & 0.5782*          & 0.7423*          \\
\textbf{text-embedding-ada-002}  & 0.3257*          & 0.4232*          & 0.5588*          & 0.3777*          & 0.4856*          & 0.6479*          \\
\midrule
\textbf{BERT-XS}    & 0.1643*          & 0.2344*          & 0.3474*          & 0.1128*          & 0.1697*          & 0.2484*          \\
\textbf{Lawformer}  & 0.3002*          & 0.3853*          & 0.4913*          & 0.3624*          & 0.4559*          & 0.5503*          \\
\textbf{SAILER}  & 0.3731*          & 0.5626*          & 0.8148*          & 0.4932*          & 0.6537*          & 0.7945*          \\

\midrule

\textbf{Caseformer (ours)} & {0.4929} & {0.6541} & \textbf{0.8323} & \textbf{0.5648} & \textbf{0.7117} & \textbf{0.8374} \\
\toprule
\end{tabular}
\end{table}

\begin{table}[h!]

\caption{The experimental results of the Caseformer retriever (dual-encoder architecture) and other baselines on LeCaRD and CAIL-LCR after finetuning. The best results are in bold. “*” denotes the result is significantly worse than Caseformer with $ p < 0.01 $ level. R@k indicates the Recall@k metric in this table. As we cannot fine-tune OpenAI models, the results for text-embedding-ada-002 in our table represent zero-shot performance.}
\label{dense_full}
  
\centering
\begin{tabular}{lccc|ccc}
\toprule
& \multicolumn{6}{c}{\textbf{Fine-tuned}} \\
\cmidrule(r){2-7}

                    & \multicolumn{3}{c}{\textbf{LeCaRD}}                 & \multicolumn{3}{c}{\textbf{CAIL-LCR}}                                   \\
                    \cmidrule(r){2-4} \cmidrule(r){5-7}
                    & {R@100}  & {R@200}  & {R@500}  & {R@100}  & {R@200}  & {R@500}  \\
                    \toprule
\textbf{BM25}       & 0.5154*          & 0.6781*          & 0.8249*           & 0.4484*          & 0.6029*         & 0.7924*          \\
\midrule
\textbf{BERT}       & 0.5292*          & 0.7067*          & 0.8506*          & 0.8299*          & 0.9189*          & 0.9701*          \\
\textbf{RoBERTa}    & 0.5825*          & 0.7169*          & 0.8692*          & 0.8361*          & 0.9232*          & 0.9741*          \\
\textbf{SEED}       & 0.5634*          & 0.7197*          & 0.8640*          & 0.8356*          & 0.9243*         & 0.9724*          \\
\textbf{Condenser}       & 0.5937*          & 0.7396*          & 0.8666*          & 0.8415*          & 0.9301*         & 0.9774*          \\
\textbf{coCondenser}       & 0.5946*          & 0.7425*          & 0.8710*          & 0.8403*          & 0.9285*         & 0.9766*          \\
\textbf{text-embedding-ada-002}  & 0.3257*          & 0.4232*          & 0.5588*          & 0.3777*          & 0.4856*          & 0.6479*          \\
\midrule
\textbf{BERT-XS}    & 0.1769*          & 0.2842*          & 0.4368*          & 0.1993*          & 0.2758*         & 0.4014*          \\
\textbf{Lawformer}  & 0.4806*          & 0.6465*          & 0.8198*          & 0.8139*          & 0.9104*         & 0.9637*          \\
\textbf{SAILER}     & 0.5937*          & 0.7310*          & 0.8714*          & 0.8404*          & 0.9265*         & 0.9786*          \\
\midrule
\textbf{Caseformer (ours)} & \textbf{0.6111} & \textbf{0.7618} & \textbf{0.8958} & \textbf{0.8479} & \textbf{0.9360} & \textbf{0.9801} \\
\toprule

\end{tabular}

\end{table}

The performance of Caseformer Retriever and other baselines on LeCaRD and CAIL-LCR datasets are reported in Table ~\ref{dense_zero}. To be specific, we adopt recall@k (R@k) as the evaluation metric to test how many cases with labels\footnote{Both LeCaRD and CAIL-LCR datasets adopt the multi-level label (0,1,2,3) to measure the relevance between the query case and candidate cases. Labels 2 and 3 indicate that the query case is strongly relevant to the candidate case. } 2 and 3  are recalled by the retrieval model in the top-k results from the whole corpus. We evaluate the retrieval performance in both zero-shot and fine-tuning settings. For the zero-shot setting, we directly use the model after pre-training without fine-tuning. For the fine-tuning setting, we adopt five-fold cross-validation to fine-tune the PLMs. Note that in the manual annotation stage of the LeCaRD and CAIL-LCR dataset, all the candidate cases are recalled by lexical matching methods including TF-IDF, BM25, and Query Likelihood~\cite{ma2021lecard}. Therefore, the annotation results have a strong bias towards lexical matching models such as BM25 in these two datasets.

Through the experimental results, we have the following observations: (1) Caseformer outperforms the traditional retrieval method BM25 in most settings on both datasets, even though the BM25 method has a strong bias on both datasets. Besides, compared with previous SOTA pre-trained language models (PLMs), Caseformer has the best performance in both zero-shot and fine-tuning settings on both datasets. These results demonstrate that Caseformer can better capture the relevance between legal cases which indicates the effectiveness of our pre-training tasks. (2) Caseformer has a more significant advantage over other PLMs in the zero-shot setting compared with the fine-tuning setting. As the zero-shot performance directly reflects the effectiveness of the pre-training tasks, this shows that Caseformer obtains more legal knowledge in the pre-training stage compared with other PLMs. (3) As lexical matching methods do not require training, BM25 performs better than most pre-trained models in the zero-shot setting. However, the performance of PLMs exceeds BM25 after fine-tuning. {Despite the outstanding performance of the text-embedding-ada-002 model in general retrieval tasks,  it falls short in the Legal Case Retrieval task compared to BM25.} (4) The pre-trained models tailored for open-domain retrieval (e.g., SEED, coCondenser) have no significant advantages over BERT and RoBERTa. The performance of legal domain PLMs (BERT-XS, Lawformer) is lower than general domain PLMs in both zero-shot and fine-tuning settings. Showing that simply inheriting the pre-training tasks in the open domain without considering the unique characteristics of the legal field has limited benefit for case retrieval.

\subsubsection{\textbf{Multilingual Pre-training}}
\begin{table}[t!]

\caption{The experimental results of Caseformer and other baselines on COLIEE 2020 Task 1. The best results are in bold. “*” denotes the result is significantly worse than Caseformer with $ p < 0.01 $ level. R@k indicates the Recall@k metric in this table.}
\label{coliee}
\centering

\begin{tabular}{cccccc}
\toprule
                                     &                            & \multicolumn{4}{c}{\textbf{COLIEE 2020 }} \\
                                     \midrule
                                     &                & \textbf{Precision@5}         & \textbf{Recall@5}     & \textbf{MRR@10}        & \textbf{MRR@50}       \\
                                     \toprule
\multirow{2}{*}{Lexical Matching} & \textbf{BM25}               & 0.4754*   & 0.5721*           & 0.7875*                 & 0.7907*                \\
                                  & \textbf{QL}                 & 0.4554*   & 0.5506*           & 0.7906*                 & 0.7934*                \\
                                     \midrule
\multirow{2}{*}{General Domain} & \textbf{BERT}                 & 0.4542*   & 0.5588*        & 0.7923*                 & 0.7948*                \\
                                & \textbf{RoBERTa}              & 0.4639*   & 0.5862*        & 0.7613*                 & 0.7635*                \\
                                     \midrule
\multirow{3}{*}{PLMs for IR}             & \textbf{Condenser}   & 0.4862*    & 0.6127*    & 0.8198*                 & 0.8213*                \\
                                         & \textbf{coCondenser} & 0.5000*    & 0.6287*      & 0.8337*                 & 0.8347*                \\
                                     & \textbf{SEED}            & 0.5308*    & 0.6952*  & 0.8683*                 & 0.8699*                \\
                                     \midrule
\multirow{3}{*}{Legal Domain}   & \textbf{Legal-BERT}           & 0.4262*    & 0.5544*      & 0.7571*                 & 0.7594*                \\
                                & \textbf{SAILER}           & \textbf{0.5446}    & 0.7152      & 0.8823                 & 0.8831                \\
                                     & \textbf{Caseformer} & 0.5440 & \textbf{0.7234}  & \textbf{0.8856}        & \textbf{0.8872} \\
                                     \toprule
                            
\end{tabular}

\end{table}

To test the generality of our approach, we apply the Caseformer retriever to English corpora. For the pre-training dataset, we adopt the Indian Legal Statute Identification (ILSI) ~\cite{paul2022lesicin} dataset. For the downstream dataset, we adopt COLIEE~\cite{rabelo2021coliee}. The experimental result of Caseformer and other baselines are shown in Table~\ref{coliee}. We can observe that Caseformer outperforms lexical matching methods, general domain PLMs, PLMs for IR, and existing legal domain PLMs in English corpora. The powerful performance of Caseformer in different languages indicates that our proposed pre-training framework is language-independent and has strong generalization ability.

\subsubsection{\textbf{Caseformer Re-ranker}}

\begin{table}[htb]

\caption[Caption for LOF]{The experimental results of the Caseformer re-ranker (cross-encoder architecture) and other baselines on LeCaRD and CAIL-LCR in the zero-shot setting. The best results are in bold. “*” denotes the result is significantly worse than Caseformer with $ p < 0.01 $ level. N@k indicates the NDCG@k metric in this table. }
\label{crosstab}
  
\centering

\begin{tabular}{lccccccc}
    \toprule
& \multicolumn{6}{c}{\textbf{Zero-shot}} \\
\cmidrule(r){2-7}
                    & \multicolumn{3}{c}{\textbf{LeCaRD}} & \multicolumn{3}{c}{\textbf{CAIL-LCR}} \\
                    \cmidrule(r){2-4} \cmidrule(l){5-7}
                    & N@5 & N@10 & N@15 & N@5 & N@10 & N@15 \\
    \midrule
    \textbf{BM25}    & 0.6843* & 0.7082* & 0.7303* & 0.7105* & 0.7303* & 0.7490* \\
    \textbf{QL}      & 0.6906* & 0.7168* & 0.7411* & 0.7389* & 0.7535* & 0.7756* \\
    \midrule
    \textbf{BERT}    & 0.6195* & 0.6293* & 0.6487* & 0.6183* & 0.6164* & 0.6259* \\
    \textbf{PROP}    & 0.5789* & 0.5982* & 0.6044* & 0.5823* & 0.5993* & 0.6090* \\
    \textbf{BERT-XS} & 0.6485* & 0.6646* & 0.6621* & 0.6631* & 0.6790* & 0.6970* \\
    \midrule
    \textbf{Caseformer (ours)} & \textbf{0.7831} & \textbf{0.8014} & \textbf{0.8065} & \textbf{0.8288} & \textbf{0.8330} & \textbf{0.8354} \\
    \bottomrule
  \end{tabular}
\end{table}

\begin{table}[htb]
\caption[Caption for LOF]{The experimental results of the Caseformer re-ranker (cross-encoder architecture) and other baselines on LeCaRD and CAIL-LCR after finetuning. The best results are in bold. “*” denotes the result is significantly worse than Caseformer with $ p < 0.01 $ level. N@k indicates the NDCG@k metric in this table. As we cannot fine-tune OpenAI models, the results for text-embedding-ada-002 in our table represent zero-shot performance.}
\label{cross-full}
\centering

  \begin{tabular}{lccccccc}
    \toprule
    & \multicolumn{6}{c}{\textbf{Fine-tuned}} \\
\cmidrule(r){2-7}
                    & \multicolumn{3}{c}{\textbf{LeCaRD}} & \multicolumn{3}{c}{\textbf{CAIL-LCR}} \\
                    \cmidrule(r){2-4} \cmidrule(l){5-7}
                    & N@5 & N@10 & N@15 & N@5 & N@10 & N@15 \\
    \midrule
    \textbf{BM25}    & 0.6843* & 0.7082* & 0.7303* & 0.7105* & 0.7303* & 0.7490* \\
    \textbf{QL}      & 0.6906* & 0.7168* & 0.7411* & 0.7389* & 0.7535* & 0.7756* \\
    \midrule
    \textbf{BERT}    & 0.7553* & 0.7697* & 0.7966* & 0.7993* & 0.8064* & 0.8085* \\
    \textbf{PROP}    & 0.7513* & 0.7563* & 0.7892* & 0.7924* & 0.8017* & 0.8032* \\
    \textbf{BERT-XS} & 0.7486* & 0.7668* & 0.7908* & 0.7828* & 0.7973* & 0.8126* \\
    \midrule
    \textbf{Caseformer (ours)} & \textbf{0.8345} & \textbf{0.8357} & \textbf{0.8394} & \textbf{0.8362} & \textbf{0.8413} & \textbf{0.8433} \\
    \bottomrule
  \end{tabular}
\end{table}

The performance of the Caseformer re-ranker and other baselines on LeCaRD and CAIL-LCR datasets are shown in Table ~\ref{crosstab}. For the zero-shot setting, we directly use the model after pre-training without fine-tuning. For the fine-tuning setting, we adopt five-fold cross-validation to fine-tune the pre-trained language models (PLMs) on each dataset. The results are shown in Table ~\ref{crosstab} and Table~\ref{tablecail19} Through the experimental results, we have the following observations:

{In case re-ranking tasks, the experimental results reveal that Caseformer achieves superior performance over all the baselines in both zero-shot and fine-tuning settings, across both datasets. After a thorough analysis, we propose two primary factors that contribute to the exceptional performance of Caseformer. }Firstly, The pre-training tasks of Caseformer are specially designed for the legal field. Compared with pre-training tasks tailored for open-domain retrieval tasks (e.g., Web search), Caseformer consider the unique characteristics of the legal field, including the structured information of legal documents and the definition of relevance in the legal field. As a result, Caseformer learns legal knowledge and relevance-matching knowledge in the pre-training stage which is useful for case retrieval tasks. Also, the proposed LJP and FDM tasks can effectively teach Caseformer to measure the case relevance based on fact description and legal judgments which resembles how legal experts annotate the relevance between case documents.

In the zero-shot setting, traditional methods such as BM25 and QL outperform general domain pre-trained models (BERT), pre-trained models tailored for IR (PROP), and legal domain pre-trained models (BERT-XS). Showing that traditional methods are still strong baselines in the zero-shot setting. An interesting observation is that, despite its huge parameter size and outstanding performance in many open-domain tasks, the text-embedding-ada-002 model provided by OpenAI is still outperformed by simple lexical methods in Legal Case Retrieval. Caseformer is the only pre-trained language model that outperforms BM25 and QL in the zero-shot setting. This indicates the potential of domain-specific pre-training tasks for legal case retrieval.

To further evaluate the ability of Caseformer, we evaluate the performance of Caseformer and other baselines on CAIL-SCM, a legal case similarity judgment dataset of the Chinese Law System. The task of CAIL-SCM is to predict which case of two candidate cases is more similar to the query case and adopt the accuracy metric to evaluate the performance. In our experiment, we tested Caseformer and three types of re-rankers including a general domain pre-trained model (BERT), a pre-trained model tailored for IR (PROP), and a legal domain pre-trained model (BERT-XS). 

The experimental result is shown in Table ~\ref{tablecail19}. We can see that Caseformer outperforms all the baselines in both zero-shot and fine-tuning settings on both test and valid datasets. Indicates that Caseformer has a strong relevance-matching ability between legal cases. Showing the effectiveness of our pre-training framework.

\begin{table}[t!]

\caption{The experimental results of Caseformer re-ranker and other baselines on CAIL-SCM. The best results are in bold.“*” denotes the result is significantly worse than Caseformer with $ p < 0.05 $ level.}
\label{tablecail19}
\centering
\begin{tabular}{llcc}
\toprule
                               &                            & \multicolumn{2}{c}{\textbf{CAIL-SCM}} \\
                               \midrule
                               &                            & \textbf{Valid Set Accuracy}    & \textbf{Test Set Accuracy}     \\
                               \toprule
\multirow{4}{*}{Zero-shot} & \textbf{BERT}              & 0.5040*             & 0.5149*            \\
                               & \textbf{BERT-XS}           & 0.5147*            & 0.5124*            \\
                               & \textbf{PROP}              & 0.5127*            & 0.5091*            \\
                               & \textbf{Caseformer} & \textbf{0.5593}   & \textbf{0.5494}   \\
                               \midrule
\multirow{4}{*}{Fine-tuned} & \textbf{BERT}              & 0.6153*            & 0.6393*            \\
                               & \textbf{BERT-XS}           & 0.6207*            & 0.6517*            \\
                               & \textbf{PROP}              & 0.6047*            & 0.6237*            \\
                               & \textbf{Caseformer} & \textbf{0.6613}   & \textbf{0.6959}  \\
                               \toprule
\end{tabular}
\end{table}

\subsection{Ablation Study}
\label{section:ablation}
\begin{table}[t!]

\caption{The experimental results of ablation study on LeCaRD in the zero-shot setting. The best results are in bold. “*” denotes the performance is significantly better than the backbone model (BERT) with $ p < 0.01 $ level.}
    
\label{table:ablation}
\centering
\begin{tabular}{lccc}
\toprule
                          & \multicolumn{3}{c}{\textbf{LeCaRD Zero-shot}}         \\
                          \midrule
                          & \textbf{NDCG@5} & \textbf{NDCG@10} & \textbf{NDCG@15} \\
                          \toprule
\textbf{only LAM}          & 0.6531*          & 0.6614*           & 0.6690*           \\
\textbf{only FDM}          & 0.7065*          & 0.7137*           & 0.7215*           \\
\textbf{only LJP}          & 0.7456*          & 0.7498*           & 0.7503*           \\
\midrule

\textbf{w/o LAM}          & 0.7542*          & 0.7623*           & 0.7725*           \\
\textbf{w/o FDM}          & 0.7513*          & 0.7537*           & 0.7582*           \\
\textbf{w/o LJP}          & 0.7112*          & 0.7263*           & 0.7426*           \\

\midrule
\textbf{Before Pre-training} & 0.6195          & 0.6293           & 0.6487           \\
\textbf{Caseformer(Full)} & \textbf{0.7831*} & \textbf{0.8014*}  & \textbf{0.8065*} \\
\toprule
\end{tabular}
\end{table}

Our proposed pre-training framework caseformer contains three pre-training tasks. To analyze the influence and effectiveness of each task, we investigate all possible combinations of loss functions for three tasks and evaluate their performance on the LeCaRD dataset under the zero-shot setting. As we use BERT to initialize our model, we also provide the result of BERT for comparison.

The experimental results are shown in Table ~\ref{table:ablation}. We have the following findings. {Firstly, each individual task contributes to the overall enhancement of the initial model's performance. Pre-training involving all three tasks leads to the highest performance while removing any task results in a decline in model performance. Secondly, removing the Legal Judgment Prediction (LJP) task leads to the most substantial performance degradation. }This shows that measuring the legal similarity between cases is important for the case retrieval task and the model obtains the ability to measure the legal similarity through the LJP task. Finally, removing the LAM task leads to a relatively small degradation which shows that modeling the legal language is useful but limited compared with the LJP task and the FDM task.

\subsection{Visual Analysis}

\begin{figure}[t!]

\centering
    \includegraphics[width=0.65\columnwidth]{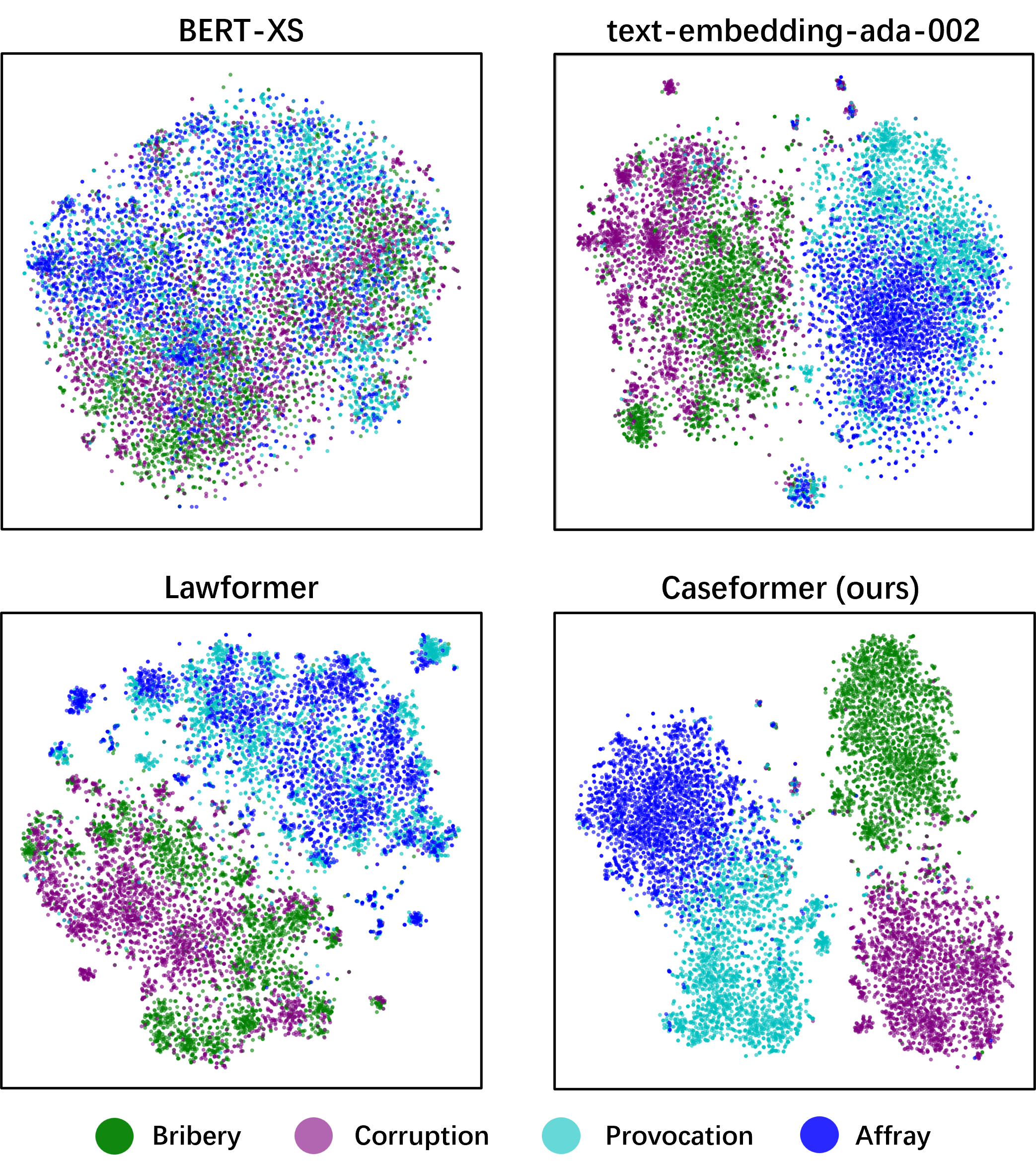}
    \caption{Visualization results of case embeddings generated by four retrieval models.}
    \label{pic:vis}
\end{figure}

To figure out the difference in the retrieval mechanism behind Caseformer and other baselines, we use t-SNE ~\cite{ van2008visualizing} as the dimension reduction method to visualize the case embeddings of different crimes. T-SNE is a nonlinear dimension reduction algorithm used to reduce the dimension of high-dimensional vectors to a lower dimension. The vectors that are close to each other will remain close after the t-SNE dimension reduction. 

Given a query case, retrieval methods aim to recall the cases that are close to the query in the embedding vector’s distance. As a result, the visualization of case embeddings can intuitively show how the model measures the relevance between cases. Specifically, we visualize the legal cases of four crimes: Bribery, Corruption, Provocation (crime of picking quarrels and provoking trouble), and Affray. We randomly select 2500 cases for each crime and visualize the embeddings generated by different retrieval models in the zero-shot setting, which is shown in Figure ~\ref{pic:vis}. Note that in Chinese criminal law, the clauses of Bribery and Corruption are two different articles but share the same \textit{category charge}\footnote{Category charge is the general name of a certain type of crime.} (the category charge of Graft and Bribery). The clauses of Provocation and Affray also share the same category charge (the category charge of Disrupting the Order of Social Administration). The cases of the same category charge are usually considered more difficult to be distinguished. 

Based on the visualization result, we have the following observations. First, BERT-XS mixes different crimes showing that BERT-XS is not able to measure the similarity of cases at the legal level. Second, text-embedding-ada-002 and Lawformer divide all cases into two categories according to the category charge. This shows that text-embedding-ada-002 and Lawformer can preliminarily measure the similarity between cases at the legal level but not accurately. They can only distinguish different categories but not different charges under the same category. Finally, Caseformer divided all cases into four categories based on the crime. This indicates that Caseformer can distinguish different crimes under the same category. Compared with existing PLMs like Lawformer and text-embedding-ada-002, Caseformer can measure the similarity between cases at the legal level more precisely. 

In summary, compared with other baselines, Caseformer can measure the legal similarity between cases more precisely in the zero-shot setting which indicates the effectiveness of our proposed pre-training framework.


\section{Conclusions and Future Works}
In this paper, we propose Caseformer, a pre-training framework tailored for legal case retrieval that achieves state-of-the-art performance in zero-shot settings and fine-tuning with full-scale data. In this framework, we propose three pre-training objectives that enable PLMs to learn massive legal knowledge and obtain relevance-matching ability in the legal field. Extensive experiments show the effectiveness of Caseformer.

There are several limitations of this paper that may need further exploration. One of the concerns is that could be significant differences in the format of legal documents in different countries. Some countries have clear formats for legal documents, so it’s easy to automatically extract information such as fact descriptions, crimes, legal provisions, etc. Meanwhile, the legal documents in some countries are not standardized, which makes it difficult to extract the above-mentioned information automatically. As the FDM and the LJP task require such structural information, those legal documents with no certain formats are not suitable for our framework. Therefore, in future work, we will explore the automatic methods to extract information such as fact descriptions, legal provisions, crimes, etc. from legal documents with no fixed format to further improve the generalizability of Caseformer.

\begin{acks}
This work is supported by the Natural Science Foundation of China (Grant No. 61732008, 62002194) and Tsinghua University Guoqiang Research Institute. 
\end{acks}

\newpage
\bibliographystyle{ACM-Reference-Format}
\bibliography{sample-base}

\end{document}